\documentclass[%
 reprint,
 amsmath,amssymb,
 aps,
]{revtex4-1}

\usepackage{graphicx}% Include figure files
\usepackage{dcolumn}% Align table columns on decimal point
\usepackage{bm}% bold math
\RequirePackage{textcomp}	%for \texttimes
\begin{document}

\preprint{APS/123-QED}

\title{Ultrawide phononic band gap for combined in-plane and out-of-plane waves}% Force line breaks with \\

\author{Osama R. Bilal}
\author{Mahmoud I. Hussein}%
\thanks{Corresponding author}
\email{mih@colorado.edu.}

\affiliation{Department of Aerospace Engineering Sciences, University of Colorado Boulder, Colorado 80309, USA}

\date{\today}% It is always \today, today,

\begin{abstract}
We consider two-dimensional phononic crystals formed from silicon and voids, and present optimized unit cell designs for (1) out-of-plane, (2) in-plane and (3) combined out-of-plane and in-plane elastic wave propagation. To feasibly search through an excessively large design space ($\sim$$10^{40}$ possible realizations) we develop a specialized genetic algorithm and utilize it in conjunction with the reduced Bloch mode expansion method for fast band structure calculations. Focusing on high-symmetry plain-strain square lattices, we report unit cell designs exhibiting record values of normalized band-gap size for all three categories. For the combined polarizations case, we reveal a design with a normalized band-gap size exceeding 60\%. 
%
%Valid PACS numbers may be entered using the \verb+\pacs{#1}+ command.
\end{abstract}

\maketitle

%\tableofcontents

Phononic crystals (PnCs) are periodic materials that exhibit distinct frequency characteristics such as the possibility of formation of \it{band gaps}. \rm Within a band gap, wave propagation is effectively prohibited. This inherent dynamical phenomenon can be utilized in a broad range of technologies at different length scales. Applications of PnCs include elastic/acoustic waveguiding \cite{khelif2004} and focusing \cite{yang2004focusing}, vibration minimization \cite{hussein2007JSV}, sound collimation \cite{christensen2007collimation}, frequency sensing \cite{El-Kady2008, Mohammadi2009}, acoustic cloaking \cite{torrent2008acoustic}, acoustic rectification \cite{Li2011diode}, opto-mechanical waves coupling in photonic devices \cite {eichenfield2009optomechanical}, thermal conductivity lowering in semiconductors \cite{Cleland2001, Landry2008, yu2010, hopkins2011reduction}, among others \cite{Olsson2009}. 

In general, it is most advantagous to have the frequency range of a band gap maximized while pulling its midpoint as low as possible in order to keep the unit cell size to a minimum. Selecting the topological distribution of the material phases inside the unit cell provides a a powerful means towards reaching this target, and this has been the focus of numerous research studies not only on PnCs but also photonic crystals (PtCs). 

The exploration for optimal unit cell designs was initiated by Cox and Dobson in 1999 \cite{Cox1999} (in the context of PtCs). The articles by Burger et al. \cite{Burger2004} and Jensen and Sigmund \cite{Jensen2011} provide a review of subsequent studies concerned with band-gap widening in PtCs. In the area of PnCs, the problem has been treated in a variety of settings and using several techniques. For example, unit cells have been optimized in one-dimension \cite{Hussein2002, hussein2006SMO} and in two-dimensions (2D) \cite{Sigmund2001, sigmund2003sdp, diaz2005SMO, Halkjaer2006, gazonas2006genetic, hussein2007optimal}, using gradient-based \cite{sigmund2003sdp, diaz2005SMO, Halkjaer2006} as well as non-gradient-based \cite{gazonas2006genetic, hussein2007optimal} techniques. Interest in band-gap size maximation has also been treated outside the scope of the unit cell dispersion problem \cite{sigmund2003sdp, rupp2007design}. In all these optimization studies the focus has been primarily on PnCs based on an infinite thickness model and a material composition consisting of two or more solid (or solid and fluid)  phases with the exception of a few investigations that considered thin-plate single-phase models \cite{diaz2005SMO, Halkjaer2006}. Recognizing the practical significance of solid-and-air PnCs with relatively large cross-sectional thickness, some studies considered the configuraton of a 2D solid matrix with periodic cylindrical voids - modeled under 2D plain-strain conditons \cite{Maldovan2006} or as a three-dimensional continuum with free surface boundary conditions \cite{Reinke2011} - and investigated the dependence of band-gap size upon the void radius. For combined out-of-plane and in-plane waves in 2D infinite-thickness PnCs formed from silicon and a square lattice of circular voids, it has been shown that the band-gap size normalized with respect to the mid-gap frequency cannot exceed 40\% \cite{Maldovan2006}. In this letter we utlize a specialized optimization algorithm in pursuit of the best unit cell solid-void distribution for the 2D plain-strain problem considering high-symmetry square lattices. We cover the cases of (1) out-of-plane, (2) in-plane and (3) combined out-of-plane and in-plane elastic wave propagation. Our search methodology is also applicable to the parallel problem of 2D PtCs optimization, where transverse-electric and transverse-magnetic waves may be considered separately \cite{Sigmund2008} or in combination \cite{Men2011}.         

% The governing equations considered to model out-of-plane and in-plane waves are provided in the next section. This is followed by the analysis of the unit cell and its representation. Details on the band structure calculation method and the optimization framework are then presented, followed by the results and conclusions.

The governing continuum equation of motion for a heterogeneous medium is 
\begin{equation}
\nabla \ldotp \mathbf{C} \colon \frac{1}{2}(\nabla \mathbf{u} + (\mathbf{u})^{\textrm{T}}) = \rho \mathbf{\ddot{u},}
\end{equation}
where $\mathbf{C}$ is the elasticity tensor, $\rho$ is the density, $\mathbf{u}$ is the displacement vector, $\mathbf{x} = \{x, y, z\}$ is the position vector, $\nabla$ is the gradient operator, and $(.)^{\textrm{T}}$ is the transpose operation. We assume the wave propagation to be confined to the x-y plane only, that is, $\partial \mathbf{u}/\partial z = 0$. As such, we have two independent sets of equations, one for out-of-plane motion and the other for in-plane motion. To obtain the band structure for a given PnC unit cell design we assume a Bloch solution to the governing equations in the form $\mathbf{u(x,k;t) = \tilde{u}(x,k)}e^{i(\mathbf{k.x}-\omega \mathbf{t})}$ where $\tilde{\mathbf{u}}$ is the Bloch displacement vector, $\mathbf{k}$ is the wave vector, $\omega$ is the frequency, $t$ is the time, and $i = \sqrt{-1}$. Due to lattice symmetry the analysis is restricted to the first Brillouin zone. We consider square lattices and furthermore impose $C_{4v}$ symmetry at the unit cell level. Subsequently design representation is needed in only a portion of the unit cell and the band structure calculation is limited to the corresponding irreducible Brillouin zone (IBZ). Furthemore, we model only the solid portion of the unit cell. The void portion is not modeled since we permit only contiguous distribution of solid material. In this manner the PnCs considered exhibit geometric periodicity (with free in-plane surfaces) and not material periodicity. In practice the voids will be either in vacuum or filled with air. Our model presents an adequate representation of both cases because the elastic waves propagating in the solid will have the dominating effect \cite{Reinke2011}. In fact, this also suggests that the results we show are practically independent of the choice of the solid material. We numerically solve the emerging eigenvalue problem using the finite element (FE) method utilizing 4-node bilinear quadrilateral elements. In addition, the reduced Bloch mode expansion (RBME) method \cite{hussein2009reduced2} is applied to the FE model to substantially speed up the band structure calculations throughout the optimization process. In the RBME implementation we use a two-point expansion. The final reported results however are based on full (non-reduced) calculations.   

%\subsection*{Unit cell representation} 
\it{Unit Cell Optimization} \rm We represent a square unit cell Y by $n$\texttimes$n$ pixels forming a binary matrix G. This matrix is then reduced in size following the underlying unit cell symmetry. Each of the pixels can be assigned to either a no-material (void) or a material (silicon), i.e., $g_{ij} \in \{0,1\}$. Throughout all the intermediate steps of the optimization process, we treat the void pixels as a highly compliant medium and this enables us to conveniently manipulate the unit cell designs. Once the optimization is complete, we assess the final designs by modeling only the silicon portion of the unit cell as described above.      

%\subsection*{Objective function } 
The objective function is formulated in terms of the size of a particular band-gap width normalized with respect to its midpoint frequency:
\begin{equation}
\label{eqn:Normalized}
f (g) = \frac{ max(min_{j=1}^{n_{k}}(\omega_{i+1}^{2}(k_{j},g))-max_{j=1}^{n_{k}}(\omega_{i}^{2}(k_{j},g)) ,0)}{(min_{j=1}^{n_{k}}(\omega_{i+1}^{2}(k_{j},g))+max_{j=1}^{n_{k}}(\omega_{i}^{2}(k_{j},g)))/2 },
\end{equation}
where $min_{j=1}^{n_{k}}(\omega_{i}^{2}(k_{j},g))$ and $max_{j=1}^{n_{k}}(\omega_{i}^{2}(k_{j},g))$ denote the minimum and maximum, respectively, of the $i^{th}$ frequency $\omega_{i}$ over the entire discrete wave vector set, $k_{j},j= 1,\dots,n_{k}$, tracing the border of the IBZ. The band gap exists only when the minimum of the ${(i+1)}^{th}$ branch is greater than the maximum of the $i^{th}$ branch; otherwise no band gap exists.

%\subsection*{GA} 
We employ a genetic algorithm (GA) to maximize $f(g)$. A GA is a nature-inspired optimization technique that mimics biological evolution. It generally starts with a pool of candidate solutions (i.e., designs) according to a certain objective (fitness) function, then applies a group of operators, namely, selection, crossover and mutation, in order to evolve to more fit designs (i.e., with higher objective function values). Compared to gradient-based methods, GAs are less likely to get trapped into local minima especially for problems with vast search spaces, hyper dimensions and large number of variables \cite{goldberg1989gas}, as is the case in our unit-cell optimization problem.  

%\subsection*{Fitness function} 
In our GA the initial population of unit-cell designs is set up to be random to avoid any initial bias that might negatively affect the search. Since it is unlikely to have a band gap at the onset, the area between the two dispersion branches of interest is subsequently used as an indicator of the fitness of the unit cell design:
\begin{equation}
\label{eqn:Area} 
Fitness = H_{o} + \phi_{1} f_{o}(g) + \phi_{2} H_{i} + \phi_{3} f_{c}(g)  , %Fitness = \phi f(g) + F,
\end{equation}
%in order to guarantee that any design that has a simultaneous band gap is selected over a design that has one type of band gaps,
where subscript $(.)_{i, o, c}$ denotes the wave type (i.e., either out-of-plane, in-plane or combined), $\phi_{1-3}$ are constants equal to $10^4, 10^8, 10^{15}$ respectively, introduced to set priorities during the evolution process, and $H_{j}$ is a step function defined as:
\begin{equation}
H_{j} =\left\{\begin{array}
{ll} 0 & \textrm {if}~ f(g) > 0 ~~~~\textrm{(band gap exists)} \\
Area & \textrm {if}~ f(g) = 0 ~~~~\textrm{(no band gap)}
\end{array}\right.
\label{eqn:area}
\end{equation}
In Eq. (\ref{eqn:area}) the \it{Area} \rm represents a measure of the ''area" in frequency-wavenumber space between the two consecutive dispersion branches of interest: 
\begin{equation}
\label{eqn:Area2} 
Area = \sum_{j=1}^{n_{k}} \big[(\omega_{i+1}^{2}(k_{j},g))-(\omega_{i}^{2}(k_{j},g))]
\end{equation}
%\subsection*{Operations}
The only condition enforced in the initial population is that adjacent pixels of the same material type appear in pairs in each row. Throughout the evolution, tournament selection and single-point crossover are the two types of operations  applied on any given pair of ''parent" unit-cell designs.  Following the unit-cell symmetry constraint, the ''offspring" mutates according to a specific probability using the rule: Select random pixel $x$; if $\sum_{r=-1}^{1} g_{x+r} > 1$, set each of the three pixels to one, otherwise, to zero. The GA terminates when no further improvement in the objective function value is noted for a prescribed number of generations. At the end of the search, the final unit-cell topology passes through a simple one-point flip local search for fine tuning and smoothening.

\it{Lead-Follow Algorithm} \rm The combined out-of-plane and in-plane optimization problem poses a challenge in setting up the objective function because it is based on two sets of independent equations. Here we adopt a unique strategy, which we refer to as a \it{lead-follow} \rm algorithm, whereby the search for a combined band gap is tackled in a two-stage fashion during the evolution process. The algorithm starts with a set of random designs and searches for a band gap for out-of-plane waves (the \it{leader}\rm) between two prescribed branches guided by Eq. (\ref{eqn:Area2}) for indication of design quality. Once the GA opens a gap, it shifts its focus to the in-plane waves (the \it{follower}\rm),  but now the branch numbers encompasing the band gap are not prescribed - they are determined by the same frequency range that spans the  band gap of the leader wave type. The objective function for the follower wave-type is set to be also the \it{Area} \rm as given by Eq. (\ref{eqn:Area2}). This lead-follow process continues until a combined band gap is found, at which point the objective function effectively switches to being the actual value of the normalized combined band gap. This process is automated through a generalized fitness function as given in Eq. (\ref{eqn:Normalized}). We note that in principle the identity of the leader and the follower can be reversed.  

%\section{Results}

\it{Results} \rm In applying the specialized GA we considered the following properties for isotropic silicon ($\lambda$ and $\mu$ denote Lame's coefficients): $\rho_{s}=2330$ Kg/m$^{3}$, $\lambda_{s}=85.502$ GPa, $\mu_{s}=72.835$ GPa, and we used a resolution of $n=32$. At this resolution, the total number of possible material distributions within the unit cell domain is $8.7$\texttimes$10^{40}$. This highlights the tremendously large search space that the GA needs to navigate through. At the end of each complete GA run, we doubled the resolution of the emerged topology to become 64\texttimes 64 pixels, and then smoothened the topology (while keeping it pixelated) by following a few simple rules. For the third case (i.e., combined waves) a splines-based solid material distribution has been subsequently generated to represent a manufacturable design.  Figure 1 presents the unit cell topologies and band structures  of the optimized unit cells for the three cases, and Table 1 lists the objective values obtained. In our results we identify a band gap number by the number of the optical branch that borders it from the top. 
\begin{table}[h]
\centering
\caption{Normalized band-gap (BG) size for the optimized unit cells}
\begin{tabular}{ccccccc} \\
\hline
Wave Type&\multicolumn{2}{c} {Out-of-Plane} &\multicolumn{1}{c} {In-Plane}& \multicolumn{2}{c} {Combined} \\
\hline
BG Number &1st&2nd&2nd& \multicolumn{2}{c} {Lowest} \\ 
%\hline
Representation &Pixels&Pixels&Pixels&Pixels&Splines\\
%\hline
Norm. BG Size &1.2270&1.1132&0.7696&0.6259&0.6021\\
\hline
%$C_{2v}$[100]&256& $1.2\times 10^{77}$&0.15\\
\end{tabular}
\label{tab:Obj_fn}
\end{table}

\begin{figure}[t]
\includegraphics[angle=0, height = 150mm,
width= 0.48\textwidth] {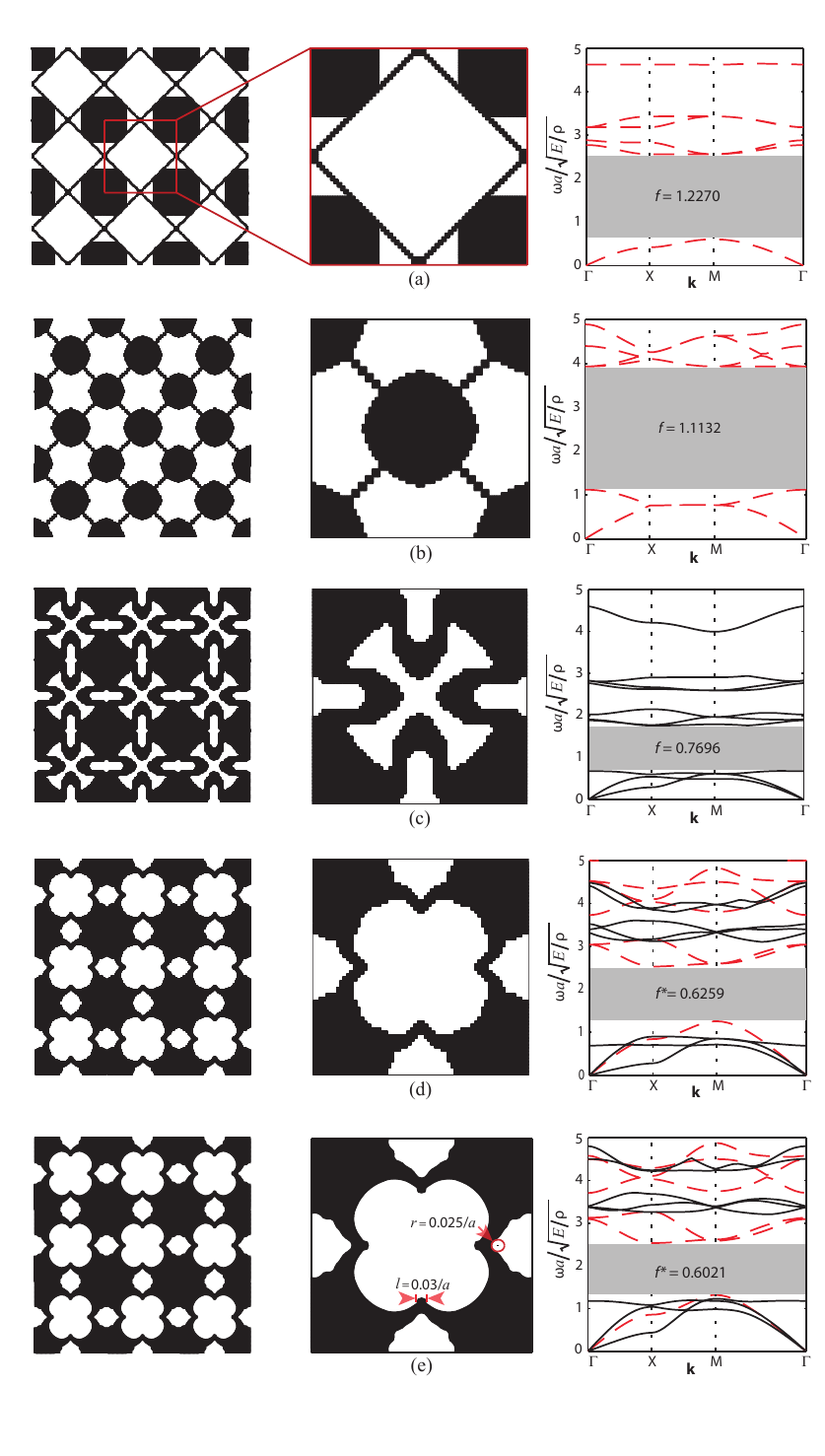}
%\end{center}
\caption{Optimized unit cell design and band structure for out-of plane waves: (a) 1st band gap, (b) 2nd band gap, in-plane waves: (c) 2nd band gap, combined out-of-plane and in-plane waves, lowest band gap: (d) pixels and (e) splines. The minimum feature size and radius of curvature are identified in (e). $f^{*}$ denotes the objective function modified to represent the normalized band-gap size for the combined waves case. All band gaps are shaded in grey.}
\label{Combined_Phononics_v3} 
\end{figure}

The optimized unit cell topologies for out-of-plane waves (Figs. 1a,b) show contiguous solid media approaching the limiting case of isolated square or circular inclusions. This limiting case represents the optimal conditions for sonic crystals which admits only pressure waves \cite{Kushwaha1996}. For the presented case, the thin connections shown are needed to support the propagation of the shear elastic waves. The optimized topology for the in-plane waves problem on the other hand show a mostly solid material with delicately shaped voids. This is consistant with the understanding in the literature that solid material with isolated voids represents the optimal conditions for band-gap opening for in-plane waves \cite{Kushwaha1996}. We note that no band gap appeared below the first optical branch due to the difficulty in preventing this branch from crossing through the acoustic branches. The optimal design for the combined case appears to be a blend (although non-intuitive in shape) among the out-of-plane and in-plane design traits. Upon appropriate size scaling to the frequency range of interest, all designs are amenable to fabrication by splines-based smoothening with minimal loss in objective value (as demonstrated in Fig. 1e). When compared to corresponding solid-void PnC unit-cell configurations reported in the literature, each of the designs shown in Fig. 1 represent a record value of normalized band-gap size in its category. 

%\section{Acknowledgement }
This research was suported by the National Science Foundation under Grant No. 0927322 (E. A. Misawa) and Grant No. 1131802 (B. M. Kramer).
%\nocite{*}
\bibliography{Bibliography_short}% Produces the bibliography via BibTeX.

\end{document}